\documentclass[aps,prc,preprint,showpacs,superscriptaddress,amsmath,amssymb,floatfix,nofootinbib]{revtex4}

\usepackage[dvips]{graphicx,color}
\usepackage{dcolumn}
\usepackage{bm}
\usepackage{here}

\def\pmb#1{\setbox0=\hbox{#1}%
  \kern-.025em\copy0\kern-\wd0 
  \kern.05em\copy0\kern-\wd0
  \kern-.025em\raise.0433em\box0 }

\def\lambdabar{\protect\@lambdabar}
\def\@lambdabar{%
\relax
\bgroup
\def\@tempa{\hbox{\raise.73\ht0
\hbox to0pt{\kern.25\wd0\vrule width.5\wd0
height.1pt depth.1pt\hss}\box0}}%
\mathchoice{\setbox0\hbox{$\displaystyle\lambda$}\@tempa}%
{\setbox0\hbox{$\textstyle\lambda$}\@tempa}%
{\setbox0\hbox{$\scriptstyle\lambda$}\@tempa}%
{\setbox0\hbox{$\scriptscriptstyle\lambda$}\@tempa}%
\egroup
}

\begin{document}

\preprint{J-PARC-TH-0266}

\title{\boldmath
Extended optimal Fermi averaging for near-recoilless $\Lambda$ production in the ($K^-$,~$\pi^-$) reaction on nuclei
}

\author{Toru~Harada}%
\email{harada@osakac.ac.jp}
\affiliation{%
Center for Physics and Mathematics,
Osaka Electro-Communication University, Neyagawa, Osaka, 572-8530, Japan
}
\affiliation{%
J-PARC Branch, KEK Theory Center, Institute of Particle and Nuclear Studies,
High Energy Accelerator Research Organization (KEK),
203-1, Shirakata, Tokai, Ibaraki, 319-1106, Japan
}
\author{Yoshiharu~Hirabayashi}%
\affiliation{%
Information Initiative Center, 
Hokkaido University, Sapporo, 060-0811, Japan
}

\date{\today}

\begin{abstract}
We propose to extend the optimal Fermi averaging procedure theoretically in order to calculate 
an in-medium $K^-n\to \pi^- \Lambda$ amplitude for the exothermic ($K^-$, $\pi^-$) reaction 
on nuclei in the framework of a distorted-wave impulse approximation, taking into account 
the local momentum transfer generated by semiclassical distorted waves for $K^-$ and $\pi^-$ mesons.
Angular distributions for the $^{12}$C($K^-$,~$\pi^-$)$^{12}_\Lambda$C reaction 
in which a momentum transfer is $q \lesssim$ 80 MeV/$c$ at $p_K=$ 800 MeV/$c$ 
in the $\pi^-$ forward direction, are estimated by applying the extended procedure.
The result shows that the calculated angular distributions are in good agreement with 
those of the data, 
and this extension is a successful prescription making it possible to describe the reaction 
cross sections in near-recoilless reactions such as ($K^-$,~$\pi^-$) in our framework.
\end{abstract}
\pacs{21.80.+a, 24.10.-i, 25.80.-e, 27.20.+n 
}

\keywords{Hypernuclei, angular distribution, DWIA, Fermi motion, local momentum
}
\maketitle


\section{Introduction}
\label{Intro}

The ($K^-$,~$\pi^-$) reaction has been an essential tool for studying spectroscopy 
in hypernuclear physics and strange particle physics \cite{Bruckner76,Bruckner78,Chrien79}. 
This reaction has played an important role in understanding hypernuclear structures 
related to the nature of $YN$ interaction, by controlling a momentum transfer 
over a wide range of $q =$ 0 to a few hundred MeV/$c$ in 
the exothermic reaction \cite{Feshbach66,Kerman71},
in comparison with endothermic reactions such as ($\pi^+$,~$K^+$) and ($e$,~$e'K^+$)
having large momentum transfers of $q \simeq$ 300--500 MeV/$c$. 

Several authors \cite{Dover79,Auerbach83,Zofka84,Bando90,Itonaga94} studied  a shell-model approach 
to $\Lambda$ hypernuclear spectroscopy in $p$-shell nuclei 
within a distorted-wave impulse approximation (DWIA), considering 
a Fermi averaging of a $K^-n \to \pi^-\Lambda$ amplitude, recoil corrections, 
and distorted waves obtained by solving the Klein-Gordon equations for $K^-$ and $\pi^-$ mesons.
The Fermi averaging treatment \cite{Rosenthal80}
may essentially affect the shape and magnitude of 
the production cross sections in the $K^-n \to \pi^-\Lambda$ reaction on nuclei  
because there appear narrow $Y^*$ resonances whose widths are smaller 
than the Fermi-motion energy of a struck nucleon in the nuclei \cite{Auerbach83}. 
These studies extract valuable information on the structure of hypernuclear states and 
the mechanism of hyperon production reactions from available experimental data at CERN, BNL, and KEK. 
The experimental studies are now in progress at J-PARC \cite{Nagae21}.

However, 
the authors \cite{Harada05,Harada06,Harada18} showed that 
the energy and angular dependence of an in-medium amplitude of 
${\overline{f}}_{\pi^-p \to K^+\Sigma^-}$ 
is significant to explain the behavior of $\Sigma^-$ production spectra for 
nuclear ($\pi^-$,~$K^+$) reactions in the DWIA, 
using the optimal Fermi averaging (OFA) procedure \cite{Harada04}, 
which provides the Fermi motion of a nucleon on the on-energy-shell 
$\pi^-p \to K^+ \Sigma^-$ reaction condition in a nucleus. 
This procedure was also applied to $\Lambda$ production via 
the ($\pi^+$,~$K^+$) reaction \cite{Harada04} and $\Xi^-$ production 
via the ($K^-$,~$K^+$) on nuclei \cite{Harada21}, 
and indicated a successful description for the $K^+$ spectra of the data.
Therefore, it seems that our OFA procedure works very well for these endothermic 
reactions characterized by large momentum transfers of $q \gtrsim$ 300--500 MeV/$c$.

Kohno and his collaborators \cite{Kohno06,Hashimoto08} also discussed the inclusive $K^+$ 
spectra via nuclear ($\pi^\pm$,~$K^+$) and ($K^-$,~$K^+$) reactions 
using the semiclassical distorted-wave (SCDW) model \cite{Luo91}.
Considering the semiclassical approximation \cite{Kawai62}, 
Luo and Kawai \cite{Luo91} showed a successful description of 
($p$, $p'x$) and ($p$, $nx$) inclusive cross sections for intermediate energy nucleon reactions 
by the SCDW model. 

When we attempt to apply the OFA procedure to a calculation for 
exothermic $\Lambda$ production reactions such as ($K^-$,~$\pi^-$) on nuclei, 
however, we realize an unavoidable difficulty that no $\Lambda$ hyperon is populated  
in hypernuclear states under a near-recoilless environment, 
e.g., $q \lesssim 80$ MeV/$c$, 
contrary to the evidence of experimental observations \cite{Bruckner76,Bruckner78,Chrien79}. 
In the OFA procedure, the on-energy-shell equation [see Eq.~(\ref{eqn:e11})] needs to satisfy 
approximately the condition as a discriminant, 
\begin{eqnarray}
 \frac{{\bm q}^2}{2\Delta m} - \Delta \omega \geq 0,
\label{eqn:e1}
\end{eqnarray}
with $\Delta \omega = \varepsilon_\Lambda({j_\Lambda}) - \varepsilon_N({j_N})$
and $\Delta m = m_\Lambda - m_N$, 
where 
$\varepsilon_\Lambda({j_\Lambda})$ and $\varepsilon_N({j_N})$
($m_\Lambda$ and $m_N$) are energies of the single-particle states (masses) 
for $\Lambda$ and $N$, respectively. 
Considering the $^{12}$C($K^-$,~$\pi^-$)$^{12}_\Lambda$C reaction at $p_K=$ 800 MeV/$c$
in the $\pi^-$ forward direction, 
we perhaps have difficulty of $q^2/2\Delta m < \Delta \omega$ due to $q \lesssim 80$ MeV/$c$.
This conjecture implies that the OFA procedure is not applicable in describing the angular distributions 
of $d\sigma/d\Omega_{\rm lab}$ in this near-recoilless ($K^-$,~$\pi^-$) reaction.

In this paper, 
we propose to extend the OFA procedure \cite{Harada04} theoretically in order 
to calculate an in-medium $K^-n\to \pi^- \Lambda$ amplitude of $\overline{f}_{K^-n\to\pi^-\Lambda}$ 
for the exothermic ($K^-$, $\pi^-$) reaction on nuclei in the framework of the DWIA, 
taking into account the local momentum transfer generated by semiclassical distorted waves 
for $K^-$ and $\pi^-$ \cite{Kawai62}.
Applying the extended OFA in the DWIA, 
we estimate the angular distributions of $d\sigma/d\Omega_{\rm lab}$ for the 
$^{12}$C($K^-$,~$\pi^-$)$^{12}_\Lambda$C reaction at $p_K=$ 800 MeV/$c$, 
in comparison with those obtained in other standard DWIA calculations.

\section{Procedure and Formulas}

\subsection{Distorted-wave impulse approximation}

We briefly mention a formulation of the angular distributions for the nuclear 
($K^-$,~$\pi^-$) reaction in the DWIA. 
Considering only the non-spin-flip processes in this reaction, 
the differential cross section for the $\Lambda$ bound state with a spin parity $J^P$ 
at the $\pi^-$ forward direction angle of $\theta_{\rm lab}$ is often written 
in the DWIA \cite{Hufner74,Dover80,Auerbach83} as (in units $\hbar = c =1$)
\begin{eqnarray}
\left({d\sigma \over d\Omega}\right)_{\rm lab}^{J^P}
&=&  \alpha \frac{1}{2J_A+1} \sum_{m_Am_B}
\biggl| \Bigl\langle {\Psi}_B 
\Big\vert\, \overline{f}_{K^-n\to\pi^-\Lambda} \nonumber\\
&\times & 
\chi^{(-)*}_{b}\left({\bm p}_{\pi},{\bm r}\right) 
\chi^{(+)}_{a}\left({\bm p}_{K},{\bm r}\right) 
\Big| \Psi_{A} \Bigr\rangle \biggr|^2,
\label{eqn:e2}
\end{eqnarray}
where $\Psi_B$ and $\Psi_A$ are wave functions of the hypernuclear final state 
and the initial state of the target nucleus, respectively. 
$\chi_{b}^{(-)}$ and $\chi_{a}^{(+)}$ are distorted waves for outgoing $\pi^-$ 
and incoming $K^-$ mesons, respectively. 
The kinematical factor $\alpha$ denotes the translation from a two-body $K^-$-nucleon 
laboratory system to a $K^-$-nucleus laboratory system \cite{Dover83}.
The energy and momentum transfers to the final state are given by
\begin{eqnarray}
&\omega = E_K-E_\pi, &\quad {\bm q} ={\bm p}_K-{\bm p}_\pi, 
\label{eqn:e3}
\end{eqnarray}
where $E_{K}=({\bm p}_{K}^2+m_{K}^2)^{1/2}$ and $E_\pi=({\bm p}_\pi^2+m_\pi^2)^{1/2}$ 
(${\bm p}_{K}$ and ${\bm p}_{\pi}$) are laboratory energies (momenta) 
of $K^-$ and $\pi^-$ in this reaction, respectively; $m_{K}$ and $m_\pi$ are
masses of $K^-$ and $\pi^-$, respectively. 
The quantity $\overline{f}_{K^-n\to\pi^-\Lambda}$ denotes the in-medium $K^-n \to \pi^- \Lambda$ 
non-spin-flip amplitude. An in-medium $K^-n \to \pi^- \Lambda$ spin-flip amplitude 
$\overline{g}_{K^-n\to\pi^-\Lambda}$ is neglected in this work
because the spin-flip part of the elementary $K^-n\to \pi^-\Lambda$ amplitude gives negligible contributions
near the forward direction in the ($K^-$,~$\pi^-$) reaction \cite{Auerbach83}.

\subsection{Local momentum transfer}

In the semiclassical approximation \cite{Kawai62}, 
the {\it local} momentum transfer in the nucleus may be defined as
\begin{eqnarray}
{\bm q}({\bm r}) 
&\equiv& 
\frac{{\rm Re}\{(-i{\bm \nabla})\chi_b^{(-)*}({\bm p}_{\pi},{\bm r})\chi_a^{(+)}({\bm p}_{K},{\bm r})\}}
{\bigl|\chi_b^{(-)*}({\bm p}_{\pi},{\bm r})\chi_a^{(+)}({\bm p}_{K},{\bm r})\bigr|}  \nonumber\\
&=& {\bm p}_K({\bm r})-{\bm p}_\pi({\bm r}),
\label{eqn:e4}
\end{eqnarray}
where ${\bm p}_K({\bm r})$ and ${\bm p}_\pi({\bm r})$ are {\it local} momenta 
for $K^-$ and $\pi^-$, respectively, which are generated by the 
semiclassical distorted waves of $\chi_a^{(+)}$ and $\chi_b^{(-)}$
that are assumed to behave as a slowly varying function of a local point ${\bm r}$ 
in the trajectory. 
We obtain these distorted waves numerically in program PIRK \cite{Eisenstein74}, 
solving the Klein-Gordon equations for the $K^-$ and $\pi^-$ mesons with 
the standard Kisslinger optical potentials, 
\begin{eqnarray}
2EU(r)=-p^2 b_0 \rho_A(r)+ b_1 {\bm \nabla} \cdot \rho_A(r) {\bm \nabla}, 
\label{eqn:e5}
\end{eqnarray}
where $\rho_A(r)$ is the nuclear density normalized to the total number of nucleons, $A$.
Considering the ($K^-$, $\pi^-$) reaction on $^{12}$C at $p_K=$ 800 MeV/$c$, 
we have $p_\pi=$ 732 MeV/$c$ for the $[(0s_{1/2})_\Lambda(0p_{3/2})_n^{-1}]_{1^-}$ state
and $p_\pi=$ 744 MeV/$c$ for the $[(0p_{3/2})_\Lambda(0p_{3/2})_n^{-1}]_{0^+,2^+}$ states 
in $^{12}_\Lambda$C.
For $K^-$, we determine the parameters of $b_0$ and $b_1$ in Eq.~(\ref{eqn:e5}), 
fitting to the data of the 800-MeV/$c$ scattering on $^{12}$C \cite{Marlow82}, 
leading to $b_0=$ $0.309+i0.498$ fm$^{3}$ and $b_1=$ 0 fm$^{3}$ at $p_K=$ 800 MeV/$c$.
For $\pi^-$ at $p_\pi=$ 732 and 744 MeV/$c$, we interpolate the values of $b_0$ and 
$b_1$ from the corresponding parameters determined by fits to the data of 
the 710- and 790-MeV/$c$ scatterings on $^{12}$C \cite{Takahashi95}; 
we have 
$b_0=$ ($-0.099+i\,0.202$) fm$^{3}$ and $b_1=$ ($-0.258+i\,0.736$) fm$^{3}$ at $p_\pi=$ 732 MeV/$c$, 
and 
$b_0=$ ($-0.095+i\,0.201$) fm$^{3}$ and $b_1=$ ($-0.228+i\,0.736$) fm$^{3}$ at $p_\pi=$ 744 MeV/$c$.

\begin{figure}[tb]
  \begin{center}
  \includegraphics[width=1.0\linewidth]{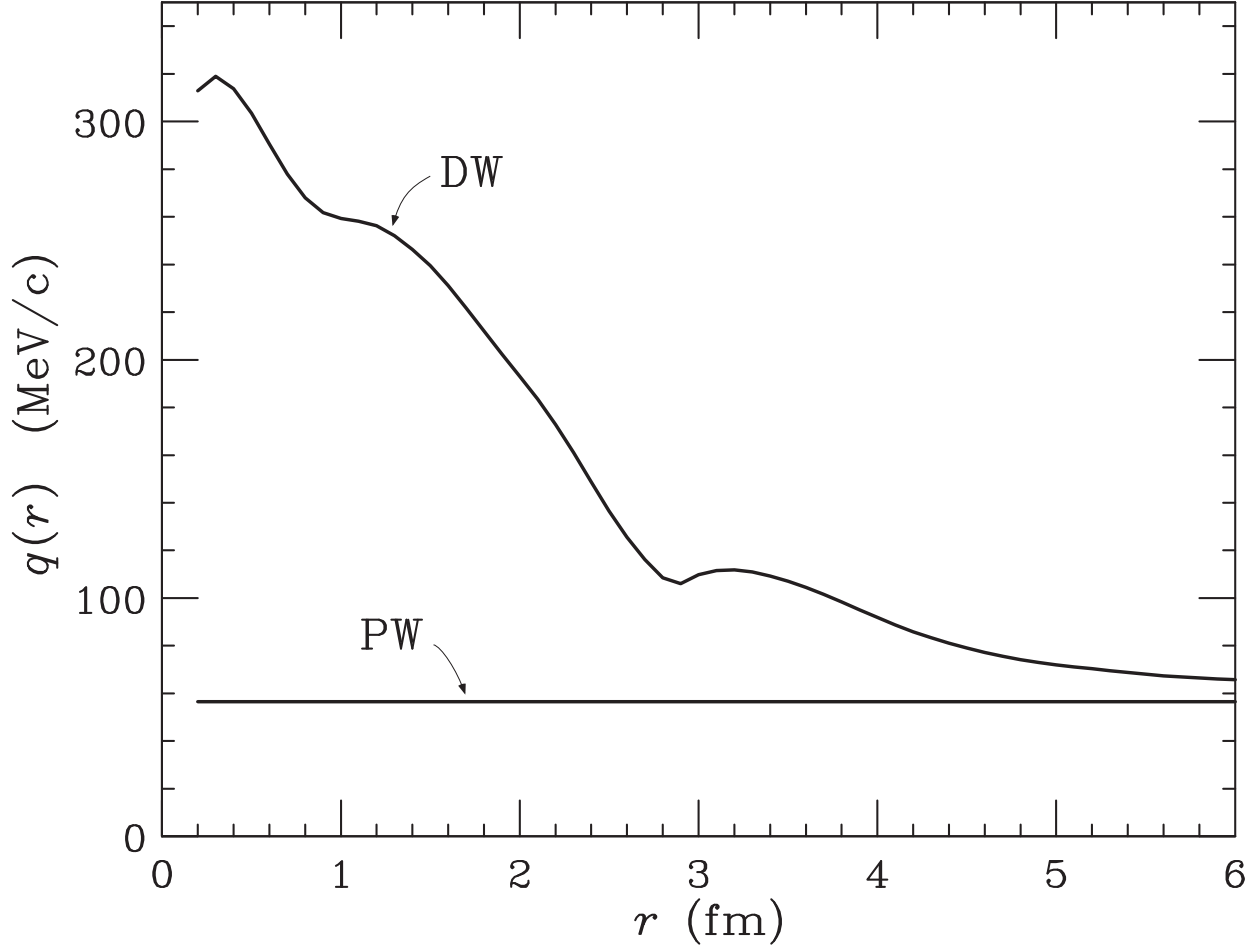}
  \end{center}
  \caption{\label{fig:1}
Magnitude of the local momentum transfers $q(r)$ for the 
$^{12}$C($K^-$,~$\pi^-$)$^{12}_\Lambda$C reaction 
at the incident $K^-$ momentum of $p_K=$ 800 MeV/$c$ and 
$\theta_{\rm lab}=$ 0$^\circ$.
The calculated values of the distorted waves (DW)
and the plane wave (PW) for the mesons are shown, as a function of the relative 
distance $r$ between the mesons and the center of the nucleus.
The asymptotic momentum transfer corresponds to $q_{K\pi}=$ 56.5 MeV/$c$.
} 
\end{figure}

\begin{figure}[tb]
  \begin{center}
  \includegraphics[width=1.0\linewidth]{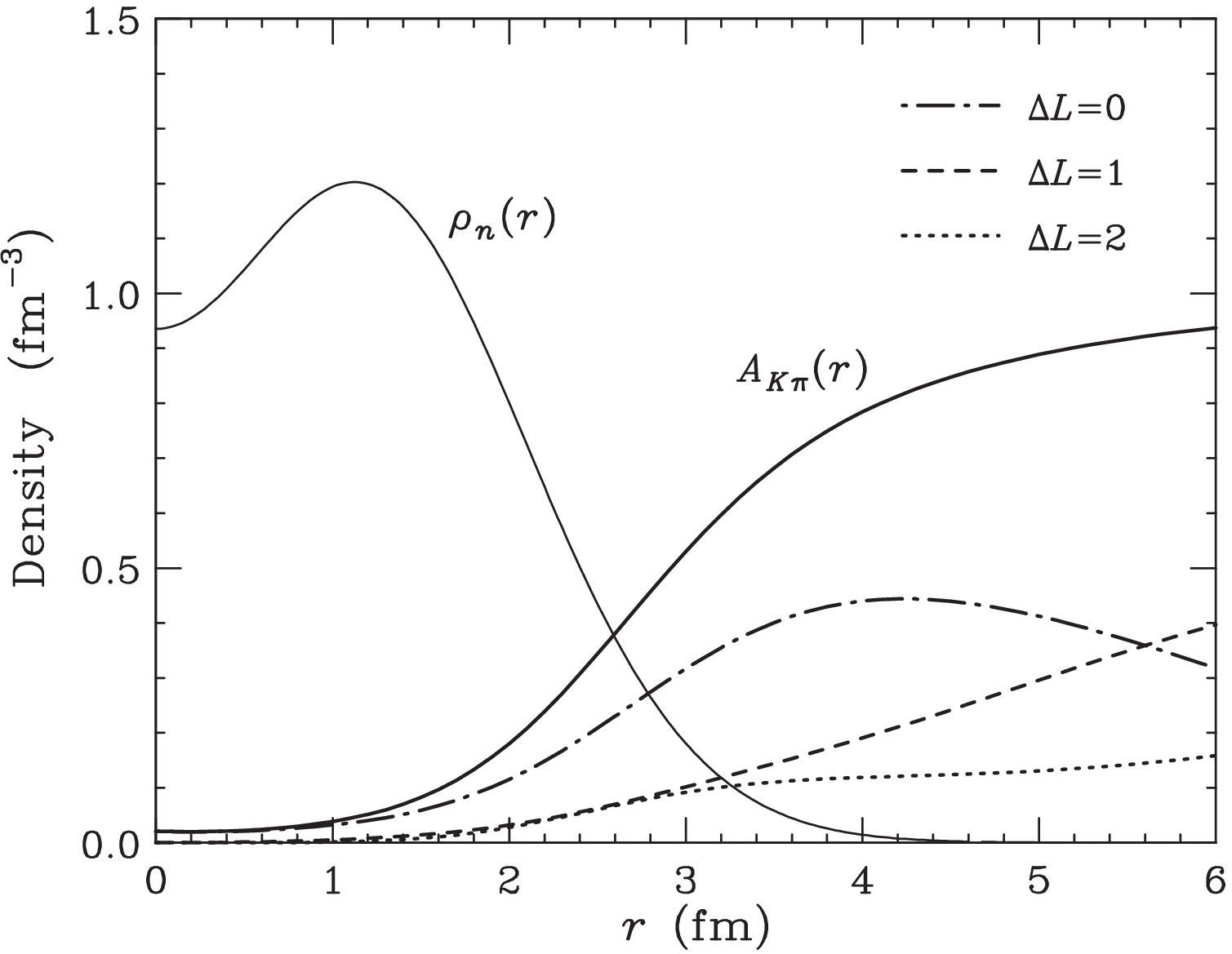}
  \end{center}
  \caption{\label{fig:2}
Meson absorption factor $A_{K\pi}(r)$ in the $^{12}$C($K^-$,~$\pi^-$)$^{12}_\Lambda$C reaction 
at the incident $K^-$ momentum of $p_K=$ 800 MeV/$c$ and $\theta_{\rm lab}=$ 0$^\circ$, 
as a function of the relative distance $r$.
Dot-dashed, dashed, and dotted curves denote the components  
of the angular momentum transfers with $\Delta L=$ 0, 1, and 2, respectively.  
The distribution of the neutron density $\rho_n(r)$ in $^{12}$C 
is also drawn. 
  } 
\end{figure}

Figure \ref{fig:1} shows the calculated values of the magnitude of the 
local momentum transfer $q(r)=|{\bm q}(r)|$ for the $^{12}$C($K^-$,~$\pi^-$)$^{12}_\Lambda$C 
reaction at $p_{K}=$ 800 MeV/$c$, 
as a function of the relative distance $r$ between the mesons and the center of the nucleus.
We find that the value of $q(r)$ amounts to about 300 MeV/$c$ at the nuclear center, 
and decreases toward the nuclear surface of $R=r_0A^{1/3}=$ 2.91 fm 
for $^{12}$C; it becomes asymptotically $q_{K\pi}=|{\bm p}_K-{\bm p}_\pi|$ 
at $|{\bm r}| \to \infty$ outside the nucleus, where $q_{K\pi}$ 
is an asymptotic momentum transfer corresponding to $q$ given in Eq.~(\ref{eqn:e3}).
This behavior is determined by the attractive (repulsive) nature of the potential 
used for $K^-$ ($\pi^-$) in Eq.~(\ref{eqn:e5}).
In the plane-wave (PW) approximation for the meson waves, the value of  $q(r)$ is 
equal to $q_{K\pi}$ as a constant.
To see clearly the effect of the local momentum transfer in the nucleus, 
we estimate the root-mean-square value of $q$ defined as 
\begin{eqnarray}
 \langle q^2 \rangle^{1/2}= 
\left[\frac{\int_0^\infty dr r^2 \rho_n(r)A_{K\pi}(r)|{\bm q}(r)|^2}
{\int_0^\infty dr r^2 \rho_n(r)A_{K\pi}(r)}
\right]^{1/2}, 
\label{eqn:e7}
\end{eqnarray}
where $\rho_n(r)$ is the neutron density in the target nucleus, 
normalized by $\int \rho_n(r)d{\bm r}=N$, and $A_{K\pi}(r)$ is a 
meson absorption factor \cite{Dover80,Matsuyama88}, 
which is given by
\begin{eqnarray}
A_{K\pi}(r)
&=&\frac{1}{4\pi}\int \left|\chi_b^{(-)*}({\bm p}_{\pi},{\bm r})
\chi_a^{(+)}({\bm p}_{K},{\bm r})\right|^2 d\Omega.  
\label{eqn:e8}
\end{eqnarray}
Figure~\ref{fig:2} shows the neutron density $\rho_n(r)$ in $^{12}$C
and the absorption factor $A_{K\pi}(r)$, as a function of the radial distance.
We use a modified harmonic oscillator model with the parameters of 
$\alpha=$ 2.234 and $b=$ 1.516 fm for $^{12}$C \cite{Vries87}, 
and the distorted waves obtained by Eq.~(\ref{eqn:e5}). 
Because the asymptotic momentum transfer is small for the near-recoilless ($K^-$,~$\pi^-$) reaction, 
the partial waves of the angular momentum transfer $\Delta L \lesssim$ 2 in $A_{K\pi}(r)$ 
contribute to the $\Lambda$ production; the component of $\Delta L=$ 0 is dominant 
inside the nucleus, whereas the components of $\Delta L=$ 1 and 2 grow toward the outside of the nucleus. 
We find $\langle q^2 \rangle^{1/2}=$ 203 MeV/$c$ at $\theta_{\rm lab}=$ 0$^\circ$, 
of which the value is sufficiently larger than $q_{K\pi}=$ 56.5 MeV/$c$.
The value of $\langle q^2 \rangle^{1/2}$ may effectively indicate 
the momentum transfer in the nucleus.  
Therefore, the local momentum transfer generated by the distorted waves 
is expected to significantly influence the $\Lambda$ production cross section 
for the near-recoilless ($K^-$,~$\pi^-$) reaction.

\subsection{Extended optimal Fermi averaging}

Following the semiclassical approximation in Ref.~\cite{Kawai62}, 
we attempt to extend the OFA procedure~\cite{Harada04} 
for the in-medium $K^-n \to \pi^-\Lambda$ amplitude of 
$\overline{f}_{K^-n\to\pi^-\Lambda}$, taking into account 
the effect of the local momentum transfer.
The extended optimal Fermi-averaged $K^-n\to \pi^-\Lambda$ $t$ matrix 
in the nucleus can be defined as 
\begin{eqnarray}
\overline{{t}^{\rm opt}}(p_K; \omega,{\bm q})
&=&
\frac{\int_0^\infty dr r^2 \rho_n(r)A_{K\pi}(r)\,{t}^{\rm opt}(p_K; \omega,{\bm q}(r))}
{\int_0^\infty dr r^2 \rho_n(r)A_{K\pi}(r)}, \nonumber\\
\label{eqn:e9}
\end{eqnarray}
where 
 ${t}^{\rm opt}(p_K; \omega,{\bm q})$ is 
the optimal Fermi-averaged $K^-n\to \pi^-\Lambda$ $t$ matrix at a point of 
($\omega$,~${\bm q}$)~\cite{Harada04}, which is given by
\begin{eqnarray}
&&{t}^{\rm \,opt}(p_K; \omega,{\bm q}) \nonumber\\
&&=\frac{
\int_0^{\pi} \sin{\theta_N}d\theta_N 
\int_{0}^{\infty} dp_{N} p_N^2 n(p_N)
\,{t}(E_{2};{\bm p}_K,{\bm p}_N)
}{
\int_0^{\pi}\sin{\theta_N}d{\theta_N} 
\int_{0}^{\infty} dp_{N} p_N^2 n(p_N)}
\Biggl|_{{\bm p}_N={\bm p}^*_N}, 
\label{eqn:e10}
\end{eqnarray}
where 
${t}(E_{2};{\bm p}_K,{\bm p}_N)$ is the two-body on-shell $t$ matrix 
for the $K^-n \to \pi^-\Lambda$ reaction in free space, 
$E_{2}=E_{K}+E_{N}$ is a total energy of the $K^- N$ system, 
and $\cos{\theta_N}= \hat{\bm p}_K\cdot\hat{\bm p}_N$;
$E_N$ and ${\bm p}_N$ are an energy and a momentum of the nucleon
in the nucleus, respectively.
The function $n(p)$ is a momentum distribution of a struck nucleon in the nucleus,
normalized by $\int n(p)d{\bm p}/(2\pi)^3=1$; 
we estimate $\langle p^2 \rangle^{1/2} \simeq$ 147 MeV/$c$, 
assuming a harmonic oscillator model 
with a size parameter $b_N=$ 1.64 fm for $^{12}$C.
The subscript ${\bm p}={\bm p}^*$ in Eq.~(\ref{eqn:e10}) means 
the integral with a constraint imposed on the valuables of ($p_N$, $\theta_N$) 
that fulfill the condition for ($p_N$, $\theta_N$) = ($p_N^*$, $\theta_N^*$) 
in an on-energy-shell momentum ${\bm p}_N^*$.
The momentum ${\bm p}_N^*$ is a solution that satisfies 
the on-energy-shell equation for a struck nucleon at the point ($\omega$, ${\bm q}$) 
in the nuclear systems, 
\begin{eqnarray}
\sqrt{({\bm p}_N^*+{\bm q})^2+m_\Lambda^2}-\sqrt{({\bm p}_N^*)^2+m_N^2}=\omega,
\label{eqn:e11}
\end{eqnarray}
where $m_\Lambda$ and $m_N$ are masses of the $\Lambda$ and the nucleon, respectively. 
Note that this procedure keeps the on-energy-shell $K^-n \to \pi^-\Lambda$ processes 
in the nucleus \cite{Gurvitz86}, so that it guarantees to take ``optimal'' values 
for $t^{\rm opt}$; 
binding effects for the nucleon and the $\Lambda$ in the nucleus 
are considered automatically when we input experimental values for the binding 
energies of the nuclear and hypernuclear states.

According to the optimal momentum approximation \cite{Gurvitz86},  
the use of the on-shell $K^-n \to \pi^-\Lambda$ $t$ matrix may be 
valid in the impulse approximation  
because the leading-order correction caused by the Fermi motion is minimized.
Therefore, the OFA procedure is 
a straightforward way of dealing with the Fermi averaging 
for the elementary reaction amplitude in the optimal momentum approximation.
Moreover, this extension in this work provides the effect of the local momentum transfers 
in the semiclassical approximation 
that meson-baryon collisions are spatially localized at a point in the nucleus 
without interfering with the collisions at different points \cite{Kawai62}.
By using the extended optimal Fermi-averaged $t$ matrix in Eq.~(\ref{eqn:e9}),  
thus, the in-medium $K^-n\to\pi^-\Lambda$ amplitude for the nucleus in Eq.~(\ref{eqn:e2}) 
is given as
\begin{eqnarray}
\overline{f}_{K^-n\to\pi^-\Lambda}
&=&
-\frac{1}{2\pi}\left(\frac{p_\pi E_\pi E_K}{\alpha p_K}\right)^{1/2}
\overline{t^{\rm \,opt}}(p_K; \omega,{\bm q}), 
\label{eqn:e12}
\end{eqnarray}
as a function of the incident $K^-$ momentum $p_K$ and the detected $\pi^-$ 
angle $\theta_{\rm lab}$ and momentum $p_\pi$ in the laboratory frame.

The on-energy-shell equation of Eq.~(\ref{eqn:e11}) has 
a solution of ${\bm p}^*_N$ under the condition of 
$q^2/2\Delta m > \Delta \omega$ in Eq.~(\ref{eqn:e1}). 
Considering the $^{12}$C($K^-$,~$\pi^-$)$^{12}_\Lambda$C reaction, 
we have $\Delta m=1115 - 940 \simeq$ 175 MeV, 
$\Delta \omega = \varepsilon_\Lambda(0p_{3/2})-\varepsilon_N(0p_{3/2}) 
\simeq -1 - (-19) =$ 18 MeV for $0^+$ and $2^+_{1,2}$ 
excited states (exc.), and 
$\Delta \omega = \varepsilon_\Lambda(0s_{1/2})-\varepsilon_N(0p_{3/2})
\simeq -11 -(-19) =$ 8 MeV for a $1^-$ ground state (g.s.). 
When $q <$ 80 MeV/$c$, it is impossible to populate a $\Lambda$ in  
the $(0p_{3/2})_\Lambda$ states in the framework of the OFA 
due to $q^2/2\Delta m < \Delta \omega$. 
When $q <$ 53 MeV/$c$, it is also impossible to populate a $\Lambda$ in  
the $(0s_{1/2})_\Lambda$ state due to $q^2/2\Delta m < \Delta \omega$.
Therefore, we expect that the extended OFA procedure will overcome the difficulty 
of $q^2/2\Delta m < \Delta \omega$
even if the near-recoilless reaction has $q \lesssim$ 80 MeV/$c$.
However, we believe that 
the standard Fermi averaging (SFA)~\cite{Auerbach83,Rosenthal80} supplies
the in-medium amplitude of $\overline{f}_{K^-n\to\pi^-\Lambda}$ by an assumption of the 
off-energy-shell components in the nuclear ($K^-$,~$\pi^-$) reaction condition.

\section{Results and discussion}

Let us consider the angular distributions for 
the $^{12}$C($K^-$,~$\pi^-$)$^{12}_\Lambda$C reaction at $p_K=$ 800 MeV/$c$ in the DWIA. 
Here we obtain the single-particle states for a neutron, using the Woods-Saxon potential~\cite{Bohr69}
with a strength parameter of $V_0^N$= $-64.8$ MeV, which is adjusted to reproduce the 
data of the charge radius of 2.46 fm \cite{Jacob66}. 
For a $\Lambda$, we calculate the single-particle states, using the Woods-Saxon potential  
with $V_0^\Lambda$= $-30.3$ MeV, $a=$ 0.60 fm, $R=$ 2.58 fm, and a spin-orbit strength of 
$V_{ls}^\Lambda=$ 2 MeV for $A=$ 12 \cite{Millener88,Gal16}, 
leading to the calculated energies of $\varepsilon_\Lambda(0s_{1/2})=$ $-11.36$ MeV, 
$\varepsilon_\Lambda(0p_{3/2})=$ $-0.60$ MeV, and 
$\varepsilon_\Lambda(0p_{1/2})=$ $-0.32$ MeV.
We perform the extended OFA for the $K^-n\to \pi^-\Lambda$ reaction, 
using the elementary amplitudes analyzed by Gopal {\sl et al}.~\cite{Gopal77}, and 
we estimate the angular distributions for the $^{12}$C($K^-$,~$\pi^-$)$^{12}_\Lambda$C 
reaction according to Eq.~(\ref{eqn:e2}).

\subsection{\boldmath
In-medium $K^-n \to \pi^-\Lambda$ differential cross sections
}

\begin{figure}[tb]
  \begin{center}
  \includegraphics[width=0.90\linewidth]{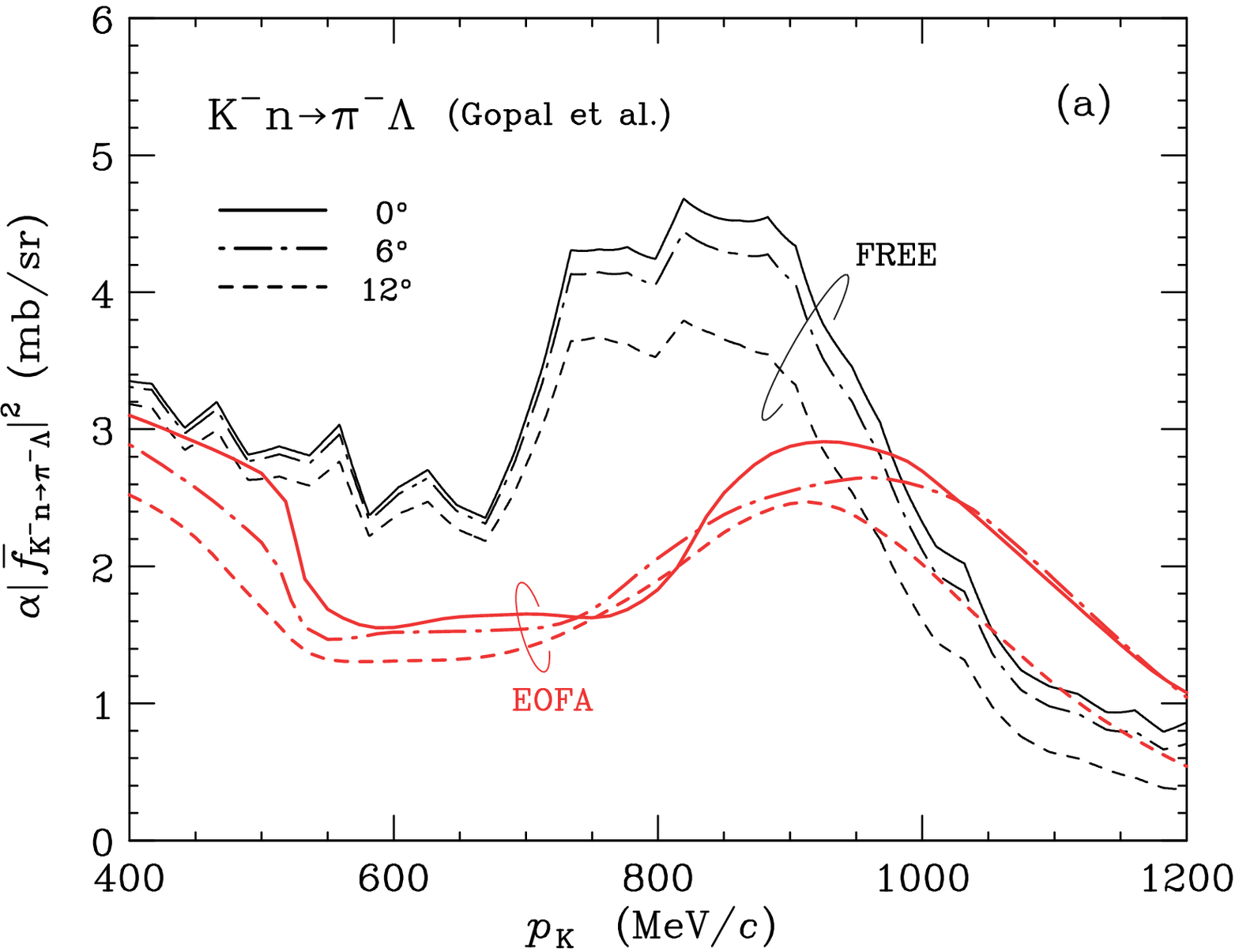}
  \includegraphics[width=0.90\linewidth]{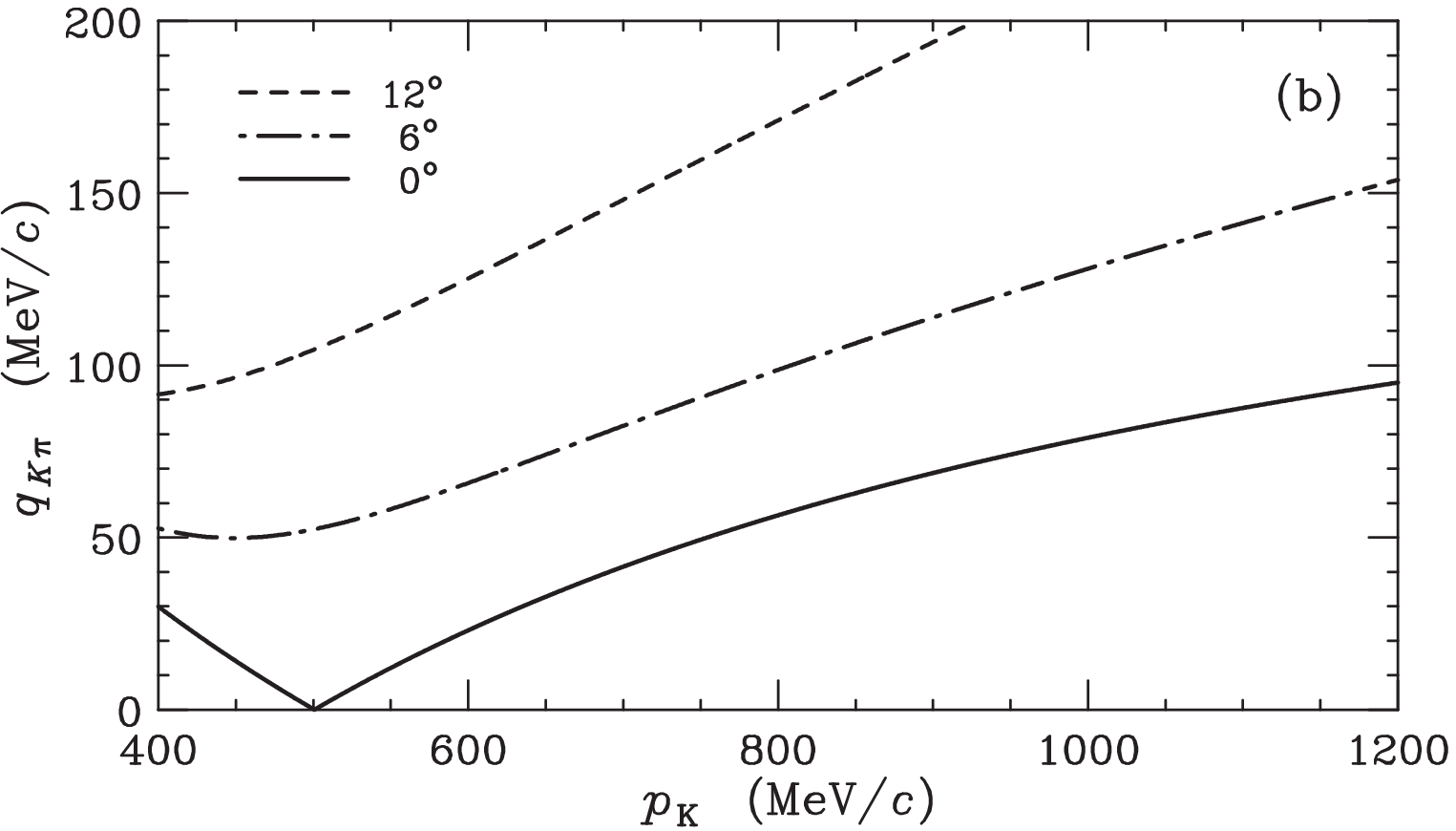}
  \end{center}
  \caption{\label{fig:3}
(a) In-medium $K^-n\to \pi^-\Lambda$ differential cross sections of 
$\alpha |\overline{f}_{K^-n\to\pi^-\Lambda}|^2$ obtained by the extended OFA (EOFA) on the $^{12}$C target, 
together with the elementary  cross sections in free space (FREE) 
including the kinematical factor $\alpha$. 
The amplitudes analyzed by Gopal {\sl et al}.~\cite{Gopal77} are used. 
Solid, dot-dashed, and dashed curves denote the values of 
$\alpha |\overline{f}_{K^-n\to\pi^-\Lambda}|^2$ for 
$\theta_{\rm lab}=$ 0$^\circ$, 6$^\circ$, and 12$^\circ$, respectively.
(b) The asymptotic momentum transfer $q_{K\pi}$
in the $^{12}$C($K^-$,~$\pi^-$)$^{12}_\Lambda$C reaction, 
as a function of $p_{K}$ in the laboratory frame.
  } 
\end{figure}

\begin{figure}[tb]
  \begin{center}
  \includegraphics[width=0.90\linewidth]{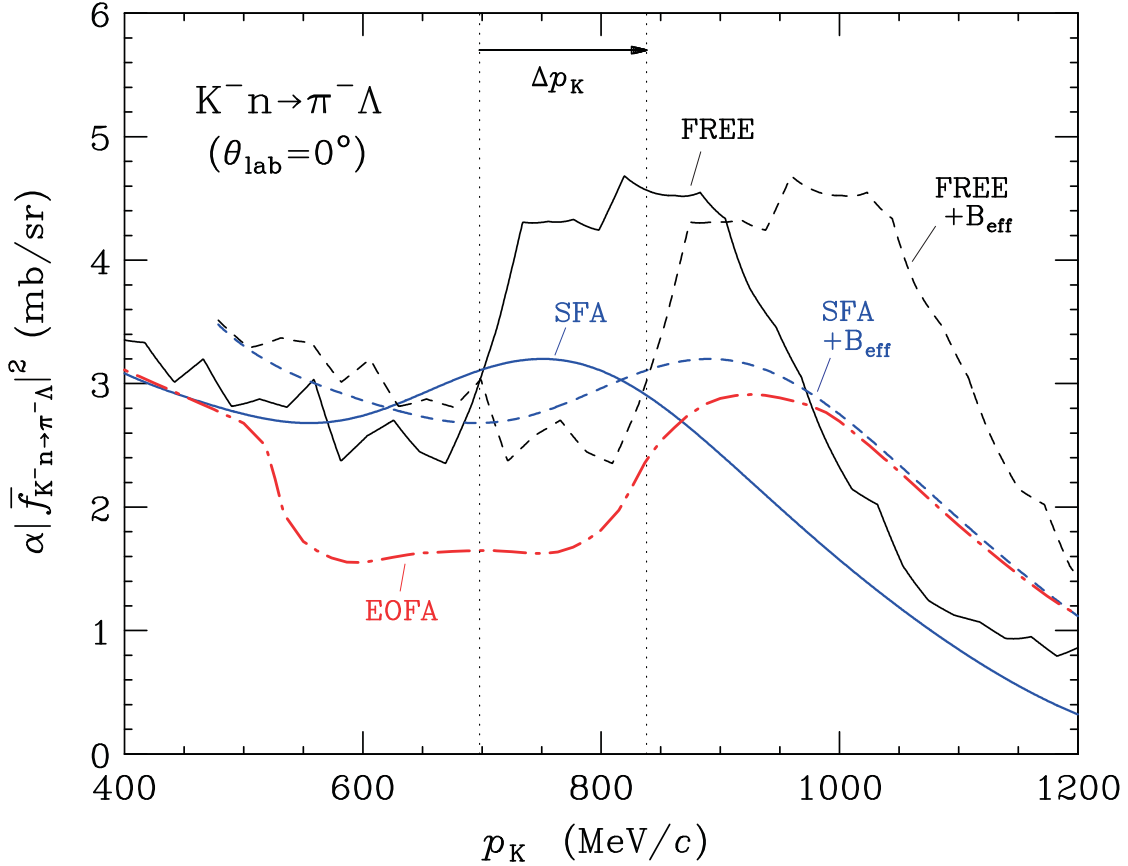}
  \end{center}
  \caption{\label{fig:4}
Comparison of the calculated results of $\alpha |\overline{f}_{K^-n\to\pi^-\Lambda}|^2$ 
obtained by the extended OFA (dot-dashed curve) with those obtained by the SFA and by the FREE 
on a $^{12}$C target 
at $\theta_{\rm lab}=$ 0$^\circ$.
Solid and dashed curves denote the values of $\alpha |\overline{f}_{K^-n\to\pi^-\Lambda}|^2$ 
without and with a 140-MeV/$c$ momentum upward shift 
of $\Delta p_K$ due to the binding effects ($B_{\rm eff}$), respectively,  
  } 
\end{figure}

Figure~\ref{fig:3} shows the calculated results of the 
in-medium $K^-n\to \pi^-\Lambda$ differential 
cross sections of $\alpha |\overline{f}_{K^-n\to\pi^-\Lambda}|^2$ on the $^{12}$C target, 
together with the asymptotic momentum transfer $q_{K\pi}$
in the $^{12}$C($K^-$,~$\pi^-$)$^{12}_\Lambda$C reaction, 
as a function of $p_{K}$ in the laboratory frame.
We find that the values of $\alpha |\overline{f}_{K^-n\to\pi^-\Lambda}|^2$ 
obtained by the extended OFA 
are reduced in the region of $p_K=$ 550--900 MeV/$c$ that corresponds to 
$q_{K\pi} \lesssim$ 80 MeV/$c$;
the peak position is located at $p_K\simeq$ 900 MeV/$c$, which seems to shift upward
in terms of the elementary cross sections in free space (FREE) \cite{Gopal77}
including the kinematical factor $\alpha$. 
These values of $\alpha |\overline{f}_{K^-n\to\pi^-\Lambda}|^2$ may be simulated by 
the extended OFA using a constant of $\langle q^2 \rangle^{1/2}\simeq$ 200 MeV/$c$ 
as an effective momentum transfer near the nuclear surface. 

We examine the behavior of $\alpha |\overline{f}_{K^-n\to\pi^-\Lambda}|^2$ 
by the extended OFA, comparing it with that obtained by the SFA \cite{Rosenthal80} 
on the $^{12}$C target at $\theta_{\rm lab}=$ 0$^\circ$, as shown in Fig.~\ref{fig:4}.
We find that the absolute values of $\alpha |\overline{f}_{K^-n\to\pi^-\Lambda}|^2$ 
by the SFA (FREE) are 1.8 (2.4) times larger than those by the extended OFA at 800 MeV/$c$
at the forward direction angles; the values by the SFA agree with the results of 
$\alpha |\langle f_L(0) \rangle|^2$ shown in Fig.~4 of Ref.~\cite{Auerbach83}.
Here we estimate the values by the SFA that includes the binding effects 
for a struck neutron via the $^{12}$C($K^-$,~$\pi^-$)$^{12}_\Lambda$C reaction 
(SFA+$B_{\rm eff}$) because the binding effects are not taken into account in the SFA. 
Such effects are roughly evaluated by a momentum shift $\Delta p_K$ that is needed to 
populate a $\Lambda$ hyperon from the $0p_{3/2}$ neutron bound in $^{12}$C, 
supplying a separation energy of 
$|\varepsilon_N(0p_{3/2})| \simeq$ $(\Delta p_K)^2/2m_K$ where 
$m_K$ is a mass of $K^-$. Thus, we have 
\begin{eqnarray}
\Delta p_K 
&=& \sqrt{2 m_K |\varepsilon_N(0p_{3/2})|} \nonumber\\
&=&
 \sqrt{2 \times 494 \times 19}\simeq 140 \,\mbox{MeV/$c$}.
\label{eqn:e13}
\end{eqnarray}
In Fig.~\ref{fig:4}, we also draw the values of 
$\alpha |\overline{f}_{K^-n\to\pi^-\Lambda}|^2$ by the SFA+$B_{\rm eff}$, 
which are shifted upward by a 140-MeV/$c$ momentum range of $\Delta p_K$ 
when the binding effects are taken into account. 
We find that these values by the SFA+$B_{\rm eff}$ 
are nearly equal to those by the extended OFA at $p_K \gtrsim$ 1000 MeV/$c$ 
because the extended OFA provides the binding effects automatically.
On the other hand, the difference between the former and the latter 
gradually becomes bigger at $p_K \lesssim$ 1000 MeV/$c$ 
due to the region of $q^2/2\Delta m < \Delta \omega$.
For the FREE, we realize that 
the position of $\alpha |\overline{f}_{K^-n\to\pi^-\Lambda}|^2$ should 
be shifted upward by the momentum range of $\Delta p_K$ 
when the binding effects are taken into account (FREE+$B_{\rm eff}$), 
as seen in Fig.~\ref{fig:4}.
Consequently, we show the validity of the extended OFA and the meaning
of the binding effects necessary to make a good description for the $K^-n \to \pi^-\Lambda$
differential cross sections at the forward direction angles.

Furthermore, we note that 
when the elementary $K^-n \to \pi^-\Lambda$ amplitudes analyzed by Zhang {\sl et al}.~\cite{Zhang13} 
are used, the calculated values of $\alpha |\overline{f}_{K^-n\to\pi^-\Lambda}|^2$ in the extended OFA 
are very similar to those analyzed by Gopal {\sl et al}.~\cite{Gopal77}.
This situation is the same as that in the SFA. 
Therefore, we believe that the dependence of $\alpha |\overline{f}_{K^-n\to\pi^-\Lambda}|^2$ 
on the elementary $K^-n\to \pi^-\Lambda$ amplitudes is relatively small in the extended OFA and the SFA.

\subsection{\boldmath
Angular distributions for the $^{12}$C($K^-$,~$\pi^-$)$^{12}_\Lambda$C reaction at 800 MeV/$c$
}

\begin{figure}[tb]
  \begin{center}
  \includegraphics[width=0.8\linewidth]{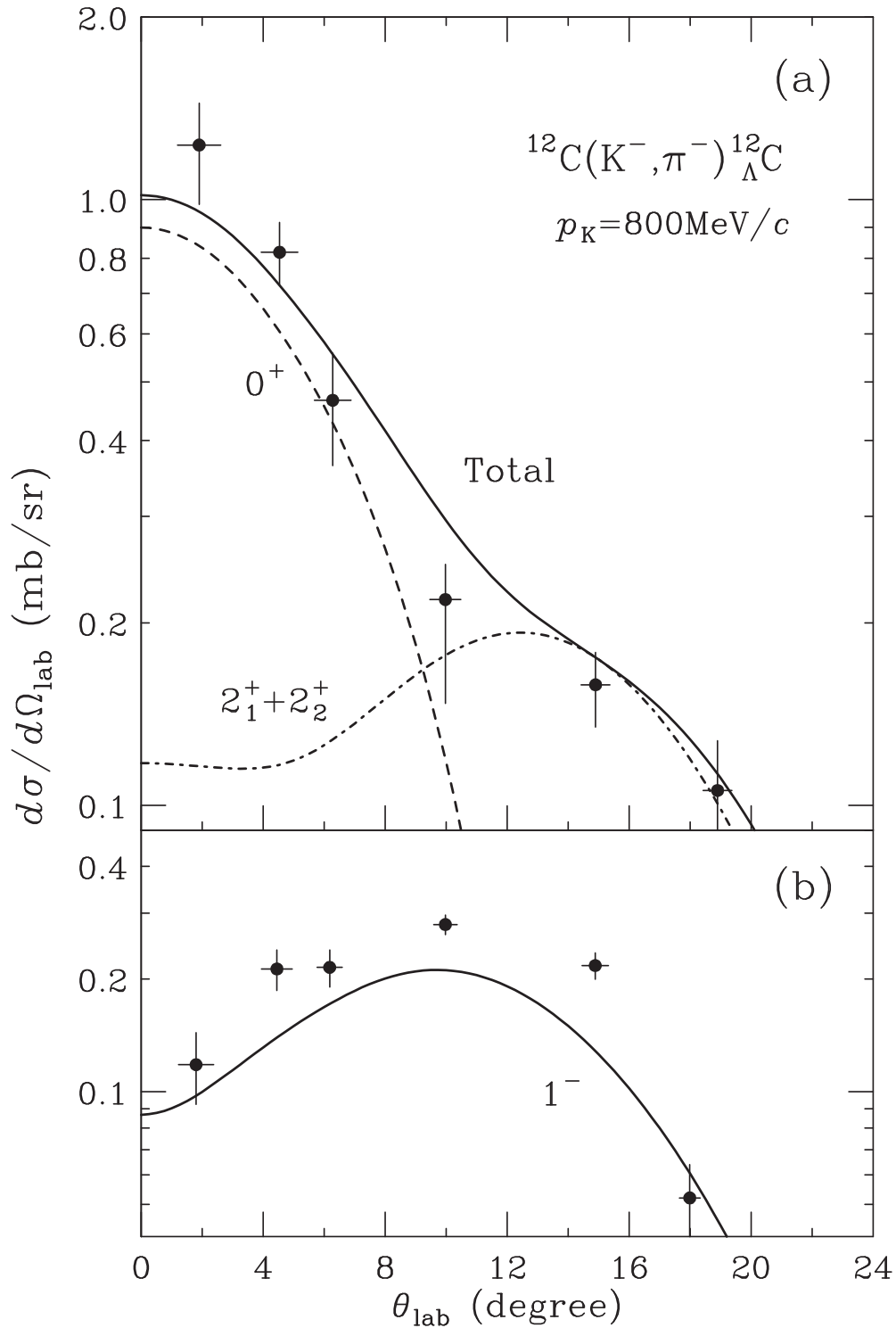}
  \end{center}
  \caption{\label{fig:5}
Calculated angular distributions of the laboratory differential cross 
sections $d\sigma/d\Omega_{\rm lab}$
for (a) the $0^+$(exc.) and $2^+_{1,2}$(exc.) states, and 
for (b) the $1^-$(g.s.) state in $^{12}_\Lambda$C 
via the $^{12}$C($K^-$,~$\pi^-$)$^{12}_\Lambda$C reaction at 
$p_K=$ 800 MeV/$c$, together with the experimental data, 
as a function of the angle of $\theta_{\rm lab}$ in the laboratory frame. 
The calculated results are obtained in the DWIA with the extended OFA.
The data are taken from Ref.~\cite{Chrien79}. 
  } 
\end{figure}

Now we estimate the angular distributions of 
the laboratory differential cross sections $d\sigma/d\Omega_{\rm lab}$ 
for $^{12}_\Lambda$C in the DWIA with the extended OFA. 
Figure~\ref{fig:5} shows the calculated values of $d\sigma/d\Omega_{\rm lab}$
for $1^-$(g.s.) and for $0^+$, $2^+_1$, and $2^+_2$(exc.) 
in $^{12}_\Lambda$C, together with the experimental data \cite{Chrien79}.
In Fig.~\ref{fig:5}(a), we display the calculated angular distributions for 
$0^+$(exc.) and $2^+_{1,2}$(exc.), 
which have the $[(0p_{3/2,1/2})_\Lambda(0p_{3/2})^{-1}_n]_{0^+}$ and 
$[(0p_{3/2,1/2})_\Lambda(0p_{3/2})^{-1}_n]_{2^+}$ configurations, respectively. 
These $\Lambda$ excited states are located near the $\Lambda$-$^{11}$C threshold
at $B^{\rm cal}_\Lambda(0^+,2^+_{1,2})=$ 0.32--0.60 MeV. 
We find that the shape and magnitude 
of the calculated sum values of $0^+$, $2^+_1$ and $2^+_2$(exc.) in the extended OFA 
are in good agreement with those of the data 
in the whole angles of $\theta_{\rm lab}=$ 0$^\circ$--20$^\circ$.
Note that the renormalization of these calculated cross sections by a factor
is not necessary to reproduce the magnitude of the data, 
in contrast with several results estimated by earlier DWIA calculations 
\cite{Dover79,Auerbach83,Bando90,Itonaga94}.
In Fig.~\ref{fig:5}(b), we display the calculated angular distribution for 
$1^-$(g.s.) having 
the $[(0s_{1/2})_\Lambda(0p_{3/2})^{-1}_n]_{1^-}$ configuration.
We find that the shape of the calculated value of $1^-$(g.s.) agrees with that 
observed in the data, 
whereas its magnitude rather underestimates at 4$^\circ$ $< \theta_{\rm lab}<$ 16$^\circ$, 
which is about 20\% smaller than that of the data. 
This discrepancy may be because more sophisticated treatments of 
nuclear wave functions are needed for more detailed comparison, e.g., 
a configuration mixing \cite{Dover79,Itonaga94}, a 2p-2h admixture, and   
other many-body effects beyond single-particle descriptions.

\begin{figure}[tb]
  \begin{center}
  \includegraphics[width=0.80\linewidth]{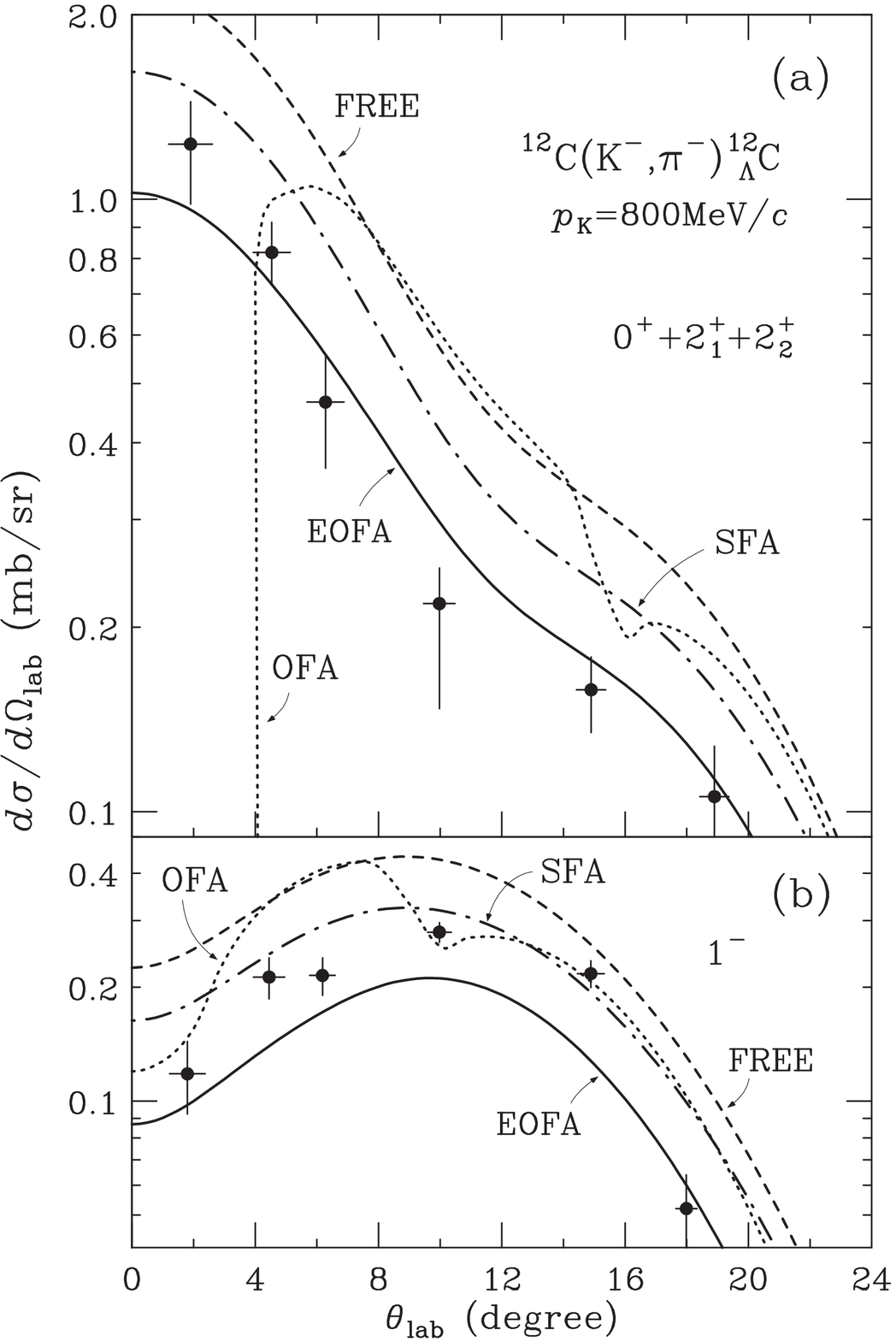}
  \end{center}
  \caption{\label{fig:6}
Comparison with the angular distributions estimated in several DWIA calculations.  
The calculated results of $d\sigma/d\Omega_{\rm lab}$ are shown for
(a) $0^+ +2^+_1 +2^+_2$(exc.) and (b) $1^-$(g.s.) in $^{12}_\Lambda$C 
via the $^{12}$C($K^-$,~$\pi^-$)$^{12}_\Lambda$C reaction at $p_K=$ 800 MeV/$c$.
Solid curves denote our results of the extended OFA (EOFA).
Dotted, dot-dashed, and dashed curves denote the values of 
the OFA omitting the effect of the local momentum transfer, the SFA, and the FREE, respectively. 
The data are taken from Ref.~\cite{Chrien79}. 
  } 
\end{figure}

Figure~\ref{fig:6} shows the comparison of our results for the extended OFA 
with those for the ``standard'' DWIA calculations. 
The quantities $\overline{f}_{K^-n\to\pi^-\Lambda}$ are estimated in the DWIA 
with the elementary amplitude in free space (FREE)~\cite{Dover79}
and 
with the SFA amplitude~\cite{Rosenthal80,Auerbach83}. 
We confirm that the magnitude of $d\sigma/d\Omega_{\rm lab}$ with the FREE amplitude is 
2.1 times as large as that of the data, and 
the magnitude with the SFA amplitude is still larger 
than that of the data by a factor of 1.6, as discussed in Ref.~\cite{Dover79}.
Therefore, it seems that the estimations of the standard DWIA are insufficient 
to explain the data, whereas these shapes of $d\sigma/d\Omega_{\rm lab}$ 
moderately agree with those of the data \cite{Dover79,Auerbach83,Bando90,Itonaga94}.
  
In the case of the OFA omitting the effect of the local momentum transfers, 
we confirm that 
the shape and magnitude of $d\sigma/d\Omega_{\rm lab}$ hardly agree with those of the data; 
no $\Lambda$ in $(0p_{3/2})_\Lambda$ is populated at $\theta_{\rm lab} \lesssim$ 4$^\circ$
due to $q^2/2\Delta m < \Delta \omega$ for $q = q_{K\pi} \simeq$ 55--80 MeV/$c$, 
whereas a $\Lambda$ in $(0s_{1/2})_\Lambda$ is populated due to 
$q^2/2\Delta m \gtrapprox \Delta \omega$.
As seen in Fig.~\ref{fig:6}(b), 
its shape behaves remarkably 
owing to the on-energy-shell condition of ($p^*_N$, $\theta^*_N$) determined from 
Eq.~(\ref{eqn:e11}) in the OFA that accompanies a Fermi averaging over a narrow width 
of $2p_N^*q/m_\Lambda$.
This result indicates that the OFA with only $q_{K\pi}$ does not work 
in the near-recoilless reaction having $q \lesssim$ 80 MeV/$c$
because of the unavoidable difficulty of $q^2/2\Delta m < \Delta \omega$
or $q^2/2\Delta m \approx \Delta \omega$.
This difficulty is overcome by the extended OFA providing 
the effect of the local momentum transfer; 
the calculated shape and magnitude of $d\sigma/d\Omega_{\rm lab}$ 
for $0^++2^+_1+2^+_2$(exc.) and $1^-$(g.s.) can explain 
those of the data without a renormalization factor quantitatively.  

Consequently, we show that the effect of the local momenta generated 
by the semiclassical distorted waves for the mesons overcomes severe 
difficulties of the previous OFA procedure in the near-recoilless 
reactions such as ($K^-$, $\pi^-$).
This result may imply the validity of the semiclassical 
picture of the localized on-energy-shell collisions by distorted waves for the mesons. 

On the other hand, it should be noticed that the endothermic 
($\pi^+$, $K^+$) reaction on nuclei satisfies the condition of 
$q(r) \lesssim q_{\pi K}$, where $q_{\pi K}$ is an 
asymptotic momentum transfer having $q_{\pi K} \simeq$ 300--500 MeV/$c$.
There is no difficulty in the OFA without handling the local momentum transfer
because the effect of the local momentum transfer 
is rather small in the nuclear ($\pi^+$,~$K^+$) reaction.
Therefore, we recognize that the OFA procedure obeying the asymptotic $q_{\pi K}$ 
works well in the ($\pi^+$, $K^+$) reaction \cite{Harada04}.

\section{Summary and conclusion}
\label{summary}

We proposed to extend the OFA procedure theoretically in order to calculate an 
in-medium $K^-n\to \pi^- \Lambda$ amplitude of $\overline{f}_{K^-n\to\pi^-\Lambda}$ 
for the exothermic ($K^-$, $\pi^-$) reaction 
on nuclei in the framework of the DWIA, 
taking into account the local momentum transfer generated by semiclassical distorted waves 
for $K^-$ and $\pi^-$.
Applying the extended OFA procedure, 
we estimated the angular distributions for the $^{12}$C($K^-$,~$\pi^-$)$^{12}_\Lambda$C reaction 
at $p_K=$ 800 MeV/$c$ under the near-recoilless condition of $q \lesssim$ 80 MeV/$c$, 
and we showed  
that the calculated angular distributions are in good agreement with those of the data. 

In conclusion, 
the extended OFA procedure provides the effect of the local momentum transfer 
generated by the meson distorted waves. 
This extension is a successful prescription making it possible 
to describe the reaction cross sections in the near-recoilless reactions such as ($K^-$,~$\pi^-$) 
in our framework.
This work may be a basis for studies clarifying the mechanism of the hadron 
production reactions on nuclei, and extracting the properties 
of a hadron-nucleus potential from the experimental data \cite{Cieply11}.

\begin{acknowledgments}
The authors thank  Professor~M.~Kawai for many valuable discussions
and comments.
This work was supported by Grants-in-Aid for
Scientific Research (KAKENHI) from the Japan Society for
the Promotion of Science: 
Scientific Research (C) (Grant No.~JP20K03954).
\end{acknowledgments}



\clearpage

\end{document}